# Supersymmetric Quantum Corrections and Poisson-Lie T-Duality

F. Assaoui* and T. Lhallabi*

The Abdus Salam International Centre for Theoretical Physics, Trieste, Italy.

**Abstract**

The quantum actions of the (4,4) supersymmetric non-linear sigma model and its dual in the Abelian case are constructed by using the background superfield method. The propagators of the quantum superfield and its dual and the gauge fixing actions of the original and dual (4,4) supersymmetric sigma models are determined. On the other hand, the BRST transformations are used to obtain the quantum dual action of the (4,4) supersymmetric non-linear sigma model in the sense of Poisson-Lie T-duality.

*

---

* Permanent address: Section of High Energy Physics, H. E. P. L, University MohammedV, Scientific Fac, Rabat, Morocco.
E-mail:  Lhallabi@fsr.ac.ma

# 1 - Introduction

Various T-duality transformations [1] connecting two seemingly different sigma models or strings backgrounds, have aroused a considerable amount of interest. The non-Abelian T-duality transformation of the isometric sigma model on a group manifold G gives non-isometric sigma model on its lie algebra [2,3]. As a result, it was not known how to perform the inverse duality transformation to get back to the original model. In order to solve this problem C. Klimcik and P. Severa [4] proposed a generalization of the Abelian and traditional non-Abelian dualities called Poisson-Lie T-duality. The main idea of this approach is to replace the requirement of isometry of sigma model with respect to some group by a weaker condition, which is the Poisson-Lie symmetry of the theory. This generalized duality is associated with two groups forming a Drinfeld double [5] and the duality transformation exchanges their roles. This approach has received further developments in a serie of works [6]. Furthermore, the Abelian and non-Abelian T-duality of the two-dimensional (4,4) supersymmetric sigma model is treated classically [7] and its Poisson-Lie T-duality is discussed [8].

On the other hand, the quantum equivalences of Poisson –Lie T-duality related sigma models were studied perturbatively in [9] and [10]. It was shown that Poisson-Lie dualizability is compatible with renormalization at 1-loop. In the present work we start by studying the quantization of the dually related (4,4) supersymmetric sigma models in the Abelian case, by using the covariant background superfield formalism [11,12]. Thereafter, we discuss the quantum equivalence of (4,4) supersymmetric sigma models related by Poisson-Lie T-duality.

The organization of this paper is as follows. In section 2, we construct the quantum actions of the (4,4) supersymmetric non-linear sigma model and its dual in the Abelian case by using the background superfield method. The propagators of the quantum superfield and its dual are obtained. Furthermore, the BRST transformations of the quantum and background superfields and their duals are given. This leads to the gauge fixing actions of the original and dual (4,4) supersymmetric sigma models. In section 3, we study the quantum Poisson-Lie T-duality of the (4,4) supersymmetric non-linear sigma model by using the BRST transformations associated with the transformation of the group elements. The generalized Cartan-Maurer equation is written in terms of the quantum supercurrents and the Poisson-Lie symmetry conditions are given. Finally, in section 4 we make concluding remarks and discuss our results.

## 2 - Quantization of the Dually Related (4,4) Supersymmetric Sigma Models in the Abelian Case

We consider a supermanifold M with metric $G_{ab}$ a = 1, . . ., d and antisymmetric tensor $B_{ab}$ which determines a supersymmetric generalized Wess-Zumino term [13]. The action of the two- dimensional (4,4) supersymmetric sigma model [7] is given by

$$S^{\sigma}_{(4,4)} = \int d\mu \left\{ (G_{ab} + B_{ab}) D^{++}\Omega^a D^{--}\Omega^b \right\} \tag{2.1}$$

where $d\mu = d^2y d^2\theta^+_+ d^2\theta^-_- du$ is the measure of the two–dimensional (4,4) analytic subspace [14] and $\Omega$ is a scalar superfield satisfying the analycity conditions [7]

$$D^+_-\Omega^a = 0 = D^-_+\Omega^a \tag{2.2}$$

with

$D^+_- = \dfrac{\partial}{\partial \theta^-_+}$ , $D^-_+ = \dfrac{\partial}{\partial \theta^+_-}$ , and $D^{\pm\pm}$ are the harmonic derivatives namely

$$\begin{aligned} D^{++} &= \partial^{++} - 2\theta^+_+ \theta^+_+ \partial_{--} \\ D^{--} &= \partial^{--} - 2\theta^-_- \theta^-_- \partial_{++} \end{aligned} \tag{2.3}$$

with

$$\partial^{\pm\pm} = U^{\pm\alpha} \dfrac{\partial}{\partial U^{\mp\alpha}}$$

In order to set up a manifestly supersymmetric covariant background field formalism for the (4,4) supersymmetric non-linear sigma model based on the parallel transport equation [11,12], we introduce the unconstrained prepotential $\lambda^a(t)$ $(0 \prec t \prec 1)$ defined along the non-geodesic curves $\Omega^a(y, \theta^+_+, \theta^-_-, U)$ [15] as:

$$\Omega^a = D^{+2}_- D^{-2}_+ \lambda^a \tag{2.4}$$

which connects the background superfield with the total superfield

$$\Omega^a = \Omega^a_{cl} + \pi^a \tag{2.5}$$

with

$$D_-^{+2} D_+^{-2} \lambda^a(0) = \Omega^a_{cl}$$
$$D_-^{+2} D_+^{-2} \lambda^a(1) = \Omega^a_{cl} + \pi^a \tag{2.6}$$

The non analytic superfield $\lambda^a$ satisfies the equation of parallel transport namely [16]:

$$\frac{d^2\lambda^a}{dt^2} + \Gamma^a_{bc}(\Omega)\frac{d\lambda^b}{dt}\frac{d\Omega^c}{dt} = 0 \tag{2.7}$$

where the solution

$$\lambda^a = \lambda^a_{cl} + \xi^a - \frac{1}{2}\Gamma^a_{bc}(\Omega_{cl})\xi^b D_-^{+2} D_+^{-2}\xi^c + ... \tag{2.8}$$

is obtained by using the initial conditions

$$\lambda^a(0) = \lambda^a_{cl}.$$
$$\left.\frac{d\lambda^a}{dt}\right|_{t=0} = \xi^a \tag{2.9}$$

$\lambda^a$ is the background prepotential and $\xi^a$ is the quantum superfield which is an unconstrained superfield. Consequently

$$\pi^a = \xi^a - \frac{1}{2}\Gamma^a_{bc}(\Omega_{cl})\xi^b D_-^{+2} D_+^{-2}\xi^c + ... \tag{2.10.a}$$

with

$$\pi^a = D_-^{+2} D_+^{-2}\pi^a. \tag{2.10.b}$$

In order to generate a covariant set of background superfield vertices in powers of the quantum superfield $\xi^a$ we insert the expansion (2.6) and (2.10) in the action (2.1). By using the developments of $G_{ab}$ and $B_{ab}$ [11] which are available in any coordinate system, the background superfield expansion of the two –dimensional (4,4) supersymmetric sigma model is given by

$$S^\sigma_{(4,4)}(\Omega) - S^\sigma_{(4,4)}(\Omega_{cl}) - \frac{\delta S^\sigma_{(4,4)}}{\delta \Omega^c}(\Omega_{cl}) D_-^{+2} D_+^{-2} \xi^c =$$

$$\int d\mu_T \{ F_{ab}(\Omega_{cl}) D^{++}\xi^a \mathbf{D}^{--} D_-^{+2} D_+^{-2} \xi^b$$

$$- \frac{1}{3} R_{abde} D_-^{+2} D_+^{-2} \xi^d D_-^{+2} D_+^{-2} \xi^e D^{++} D_-^{+2} D_+^{-2} \xi^a D^{--}\xi^b$$

$$- \frac{1}{3} R_{abde} \xi^d D_-^{+2} D_+^{-2} \xi^e D^{++}\Omega_{cl}^a D^{--}\Omega_{cl}^b$$

$$- \frac{1}{3} R_{abde} D_-^{+2} D_+^{-2} \xi^d D_-^{+2} D_+^{-2} \xi^e D^{++}\xi^a D^{--}\Omega_{cl}^b + ...\}$$

(2.11)

with $d\mu_T = d\mu\, d^2\theta_-^+ d^2\theta_+^-$, $F_{ab} = G_{ab} + B_{ab}$ and $R_{abde} = R^G_{abde} + R^B_{abde}$ (2.12)

which is the curvature tensor of the supermanifold M. $\mathbf{D}^{--}\xi^b$ is the supercovariant derivative defined by:

$$\mathbf{D}^{--}\xi^b = D^{--}\xi^b + \Gamma^b_{nm} \xi^n D^{--}\Omega_{cl}^m \quad (2.13)$$

From the action (2.11) we see that the coefficients at all orders in the quantum non analytic superfield $\xi^a$ are constructed from geometrical tensors that are functions of the background analytic superfield $\Omega_{cl}$. However, the formulation of manifestly supersymmetric Feynman rules in terms of the analytic superfield $\xi^a$ is not quite suitable in diagrammatic calculations. For this reason we introduce n-bein $e^i_a(\Omega_{cl})$ where n is the dimension of the supermanifold:

$$e^i_a(\Omega_{cl}) e_{ib}(\Omega_{cl}) = F_{ab}(\Omega_{cl}) \quad (2.14)$$

by defining

$$\xi^i = e^i_a(\Omega_{cl})\xi^a \quad (2.15)$$

the kinetic term in (2.11) becomes

$$\int d\mu_T\, D^{++}\xi^i \mathbf{D}^{--} D_-^{+2} D_+^{-2} \xi^i \quad (2.16)$$

with

$$\mathbf{D}^{--}\xi^i = D^{--}\xi^i + V_j^{--i}\xi^j$$
$$V_j^{--i}(\Omega_{cl}) \equiv V^i_{mj} D^{--}\Omega_{cl}^m \quad (2.17)$$

where $V_j^{--i}$ is the superconnection on the supermanifold. Consequently the propagator of the quantum superfield $\xi^i$ is as follows:

$$G(z_1, u_1, z_2, u_2) = \langle \xi^i(z_1, u_1) \xi^j(z_2, u_2) \rangle$$
$$= -\delta^{ij}(D_1^{++} D_1^{--} D_{-1}^{+2} D_{+1}^{-2})^{-1} \delta(z_1 - z_2) \delta(u_1, u_2) \quad (2.18)$$

with $\delta(z_1 - z_2)$ is the full (4,4) harmonic superspace $\delta$ function and $\delta(u_1, u_2)$ is the harmonic $\delta$ function [17]. It is clear that the Feynman rules constructed from the expanded action (2.11) lead to manifestly covariant quantum corrections written as integrals on the full harmonic (4,4) superspace. The divergences can be removed by counterterms namely:

$$\Delta S_\sigma^{(4,4)} = \int d\mu_T \, \Delta L \quad (2.19)$$

where the lagrangian counterterm $\Delta L$ is a scalar function of the background superfield of dimension (-2) and zero Lorentz and Cartan-Weyl charges.

The above procedure is adequate to establish a covariant background superfield expansion but the use of the unconstrained superfield $\xi^a$ means that the action (2.11) has a quantum gauge invariance which must be gauge fixed. In fact, the gauge transformations of quantum and background superfields are given by

$$\delta \xi^a = \vartheta^a$$
$$\delta \Omega_{cl} = 0 \quad (2.20)$$

where $\vartheta^a$ is a superparameter. The covariant gauge fixing term is obtained by introducing a BRST operator [14] namely

$$s\xi^a = C^a, \quad s\Omega_{cl} = 0$$
$$sC^a = 0 \quad (2.21)$$

with $C^a$ an anticommuting ghost superfield associated to the superparameter $\vartheta^a$. Furthermore, we introduce an antighost $C'^a$ and a commuting $b^a$ superfields such that:

$$sC'^a = b^a$$
$$sb^a = 0, \quad (2.22)$$

which allow to obtain a (4,4) supersymmetric gauge fixing term with ghost number zero. By using dimensional arguments and postulating BRST invariance of the action $S^{\sigma}_{(4,4)}$, the s-invariant gauge fixing action is given by

$$\tilde{S}^{\sigma}_{(4,4)GF} = \int d\mu_T \ s\{F_{ab}(\Omega_{cl})(C'^a D^{++}D^{--}\xi^b + \alpha C'^a \partial_{--}\partial_{++}\Omega^b_{cl}) \\ + \beta R_{abed} D^{++}\Omega^a_{cl} C'^b D^{++}D^{--}\xi^e D^{--}\Omega^d_{cl} \} \qquad (2.23)$$

where $\alpha$ and $\beta$ are coupling constants. The use of (2.21), (2.22) and (2.14) in the expansion of (2.23) leads to

$$S^{\sigma}_{(4,4)GF} = \int d\mu_T \ \{b^i D^{++}D^{--}\xi_i + C'^i D^{++}D^{--}C_i + \alpha b^i \partial_{++}\partial_{--}\Omega_{cli} \\ + \beta R_{aijd} D^{++}\Omega^a_{cl} b^i D^{++}D^{--}\xi^j D^{--}\Omega^d_{cl} \\ + \beta R_{aijd} D^{++}\Omega^a_{cl} C'^i D^{++}D^{--}C^j D^{--}\Omega^d_{cl} \} \qquad (2.24)$$

Eliminating the auxiliary superfield b by using its equation of motion we obtain the following constraint

$$D^{++}D^{--}\xi_i + \alpha \partial_{++}\partial_{--}\Omega_{cli} + \beta R_{aijd} D^{++}\Omega^a_{cl} D^{++}D^{--}\xi^j D^{--}\Omega^d_{cl} = 0 \qquad (2.25)$$

Consequently the gauge fixing action (2.24) becomes:

$$S^{\sigma}_{(4,4)GF} = \int d\mu_T \ \{C'^i D^{++}D^{--}C_i + \beta R_{aijd} D^{++}\Omega^a_{cl} C'^i D^{++}D^{--}C^j D^{--}\Omega^d_{cl} \} \qquad (2.26)$$

Let us now give the quantization of the dual (4,4) supersymmetric non linear sigma model which is given by [7]:

$$\tilde{S}^{\sigma}_{(4,4)} = \int d\mu \ \{(\tilde{G}_{ab} + \tilde{B}_{ab}) D^{++}Y^a D^{--}Y^b\} \qquad (2.27)$$

with

$$\begin{aligned}
Y^a &= \{\tilde{\varphi} \ , \ \Omega^i\} \quad and \\
\tilde{G}_{00} &= G^{-1}_{00} \qquad \tilde{G}_{0i} = G^{-1}_{00}B_{0i} \qquad \tilde{G}_{i0} = G^{-1}_{00}B_{0i} \\
\tilde{B}_{00} &= 0 \qquad \tilde{B}_{0i} = G^{-1}_{00}G_{0i} \qquad \tilde{B}_{i0} = -G^{-1}_{00}G_{0i} \\
\tilde{G}_{ij} &= G_{ij} - G^{-1}_{00}[G_{0i}G_{j0} + B_{0i}B_{j0}] \\
\tilde{B}_{ij} &= B_{ij} + G^{-1}_{00}[G_{0i}B_{j0} + G_{0j}B_{0i}]
\end{aligned} \qquad (2.28)$$

As previously, the background superfield expansion of the dual (4,4) supersymmetric sigma model action is given by using the developments of the dual metric tensor $\tilde{G}_{ab}$ and the dual antisymmetric tensor $\tilde{B}_{ab}$ which are available in any coordinate system [11]

$$\tilde{S}^{\sigma}_{(4,4)}(Y) - \tilde{S}^{\sigma}_{(4,4)}(Y_{cl}) - \frac{\delta \tilde{S}^{\sigma}_{(4,4)}}{\delta Y^b}(Y_{cl}) D_-^{+2} D_+^{-2} \tilde{\xi}^b =$$

$$\int d\mu_T \{ \tilde{F}_{ab}(Y_{cl}) D^{++}\tilde{\xi}^a \mathbf{D}^{--} D_-^{+2} D_+^{-2} \tilde{\xi}^b$$

$$-\frac{1}{3} \tilde{R}_{abde} D_-^{+2} D_+^{-2} \tilde{\xi}^d D_-^{+2} D_+^{-2} \tilde{\xi}^e D^{++} D_-^{+2} D_+^{-2} \tilde{\xi}^a D^{--} \tilde{\xi}^b$$

$$-\frac{1}{3} \tilde{R}_{abde} \tilde{\xi}^d D_-^{+2} D_+^{-2} \tilde{\xi}^e D^{++} Y_{cl}^a D^{--} Y_{cl}^b$$

$$-\frac{1}{3} \tilde{R}_{abde} D_-^{+2} D_+^{-2} \tilde{\xi}^d D_-^{+2} D_+^{-2} \tilde{\xi}^e D^{++} \tilde{\xi}^a D^{--} Y_{cl}^b + ...\}$$

(2.29)

with

$$\tilde{F}_{ab} = \tilde{G}_{ab} + \tilde{B}_{ab} \tag{2.30}$$

$$\tilde{R}_{abde} = R^{\tilde{G}}_{abde} + R^{\tilde{B}}_{abde} \tag{2.31}$$

and

$$\tilde{\Omega}^a_{cl} = Y^a_{cl} = \{\tilde{\varphi}, \Omega^i_{cl}\}$$
$$\tilde{\Omega}^a = Y^a = \{\tilde{\varphi} + \tilde{\pi}^0, \Omega^i_{cl} + \pi^i\} = Y^a_{cl} + \tilde{\pi}^a$$

where

$$\tilde{\pi}^a = D_-^{+2} D_+^{-2} \{\tilde{\xi}^a - \frac{1}{2} \tilde{\Gamma}^a_{bc}(\tilde{\Omega}_{cl}) \tilde{\xi}^b D_-^{+2} D_+^{-2} \tilde{\xi}^c + ...\}$$

$$\tilde{\xi}^a = \{\tilde{\xi}^0, \xi^i\}$$

By insering the Buscher's formulas (2.28) in the definition of the curvature tensor (2.31) we obtain the following relations

$$\begin{aligned}
R^{\tilde{G}}_{00de} &= -(G_{oo})^{-2} R^{G}_{00de} \\
R^{\tilde{G}}_{0ide} &= -(G_{oo})^{-2} \left[ B_{0i} R^{G}_{00de} - G_{oo} R^{B}_{0ide} \right] \\
R^{\tilde{G}}_{ijde} &= R^{G}_{ijde} - (G_{oo})^{-1} \left[ G_{0j} R^{G}_{0ide} + G_{oi} R^{G}_{0jde} - B_{0j} R^{B}_{0ide} - B_{oi} R^{B}_{0jde} \right] \\
&\quad + (G_{oo})^{-2} \left[ G_{0i} G_{0j} - B_{oi} B_{oj} \right] R^{G}_{00de}
\end{aligned} \qquad (2.32)$$

and

$$\begin{aligned}
R^{\tilde{B}}_{00de} &= 0, \qquad R^{\tilde{B}}_{0ide} = -(G_{oo})^{-2} \left[ G_{0i} R^{G}_{00de} - G_{oo} R^{G}_{0ide} \right] \\
R^{\tilde{B}}_{ijde} &= R^{B}_{ijde} - (G_{oo})^{-1} \left[ B_{0j} R^{G}_{0ide} + G_{oi} R^{B}_{0jde} - B_{0i} R^{G}_{0jde} - G_{oj} R^{B}_{0ide} \right] \\
&\quad + (G_{oo})^{-2} \left[ G_{0i} B_{0j} - B_{oi} G_{oj} \right] R^{G}_{00de}
\end{aligned} \qquad (2.33)$$

We note that these relations are similar to that of the ordinary case given in Ref [18] by imposing a quantum duality condition. It is clear that the Feynman rules constructed from the expanded dual action (2.29) will yield manifestly covariant quantum corrections written as integrals over the full harmonic (4,4) superspace. The divergences can be removed by conterterms which are integrals in the (4,4) harmonic superspace of globally defined scalar functions of the dual background superfield

$$\Delta \tilde{S}^{\sigma}_{(4,4)} = \int d\mu_T \, \Delta \tilde{L} \qquad (2.34)$$

On the other hand, the propagators of the dual quantum superfield $\tilde{\xi}^a$ are not standard as in the original theory. This can be surmoved by introducing the n-dual bein $\tilde{e}^k_a(Y_{cl})$ namely

$$\tilde{e}^k_a(Y_{cl}) \, \tilde{e}_{kb}(Y_{cl}) = \tilde{F}_{ab}(Y_{cl}) \qquad (2.35)$$

by defining

$$\tilde{\xi}^k = \tilde{e}^k_a(Y_{cl}) \tilde{\xi}^a \qquad (2.36)$$

The insertion of (2.35) and (2.36) in the action (2.29) leads to the following kinetic term

$$\int d\mu_T D^{++} \tilde{\xi}^k \mathbf{D}^{--} D^{+2}_{-} D^{-2}_{+} \tilde{\xi}^k \qquad (2.37)$$

with

$$\mathbf{D}^{--}\tilde{\xi}^k = D^{--}\tilde{\xi}^k + \tilde{V}_l^{--k}\tilde{\xi}^l$$
$$\tilde{V}_l^{--k}(Y_{cl}) \equiv \tilde{V}_{nl}^k D^{--}Y_{cl}^n \qquad (2.38)$$

where $\tilde{V}_l^{--k}$ is the dual superconnection of the supermanifold. Therefore, the propagator of the dual quantum superfield $\tilde{\xi}^k$ is given by

$$\tilde{G}(z_1, u_1, z_2, u_2) = \langle \tilde{\xi}^k(z_1, u_1) \tilde{\xi}^l(z_2, u_2) \rangle$$
$$= -\delta^{kl}(D_1^{++} D_1^{--} D_{-1}^{+2} D_{+1}^{-2})^{-1} \delta(z_1 - z_2) \delta(u_1, u_2) \qquad (2.39)$$

Furthermore, the dual action (2.29) has a quantum gauge invariance, which must be gauge fixed leading to Faddeev-Popov ghosts in the usual way. In fact, the gauge transformations of quantum and background dual superfields are given by

$$\delta \tilde{\xi}^a = \tilde{\vartheta}^a$$
$$\delta(Y_{cl}^a) = 0 \qquad (2.40)$$

with $\tilde{\vartheta}^a$ a superparameter, and their corresponding BRST transformations are

$$s\tilde{\xi}^a = \tilde{C}^a \,, \quad sY_{cl}^a = 0$$
$$s\tilde{C}^a = 0 \qquad (2.41)$$

$\tilde{C}$ is a dual ghost superfield associated with the superparameter $\tilde{\vartheta}^a$. However the s-invariant gauge fixing dual action is given by

$$\tilde{S}_{(4,4)GF}^{\sigma} = \int d\mu_T \, s\{\tilde{F}_{ab}(Y_{cl})(\tilde{C}'^a D^{++}D^{--}\tilde{\xi}^b + \alpha'\tilde{C}'^a \partial_{--}\partial_{++}Y_{cl}^b) + \beta' \tilde{R}_{abed} D^{++}Y_{cl}^a C'^b D^{++}D^{--}\tilde{\xi}^e D^{--}Y_{cl}^d \} \qquad (2.42)$$

where $\alpha'$ and $\beta'$ are coupling constants and

$$s\tilde{C}'^a = \tilde{b}^a$$
$$s\tilde{b}^a = 0 \qquad (2.43)$$

By using (2.35), (2.41) and (2.43) in the dual gauge fixing action (2.42) we obtain

$$\tilde{S}^{\sigma}_{(4,4)GF} = \int d\mu_T \left\{ \tilde{b}^k D^{++}D^{--}\tilde{\xi}_k + \tilde{C}'^k D^{++}D^{--}\tilde{C}_k + \alpha'\tilde{b}^{li}\partial_{++}\partial_{--}Y_{cl\,k} \right.$$
$$+ \beta' \tilde{R}_{akld} D^{++}Y^a_{cl} \tilde{b}^k D^{++}D^{--}\tilde{\xi}^l D^{--}Y^d_{cl}$$
$$\left. + \beta' \tilde{R}_{akld} D^{++}Y^a_{cl} \tilde{C}'^k D^{++}D^{--}\tilde{C}^l D^{--}Y^d_{cl} \right\} \qquad (2.44)$$

Eliminating the auxiliary dual superfield $\tilde{b}$ by using its equation of motion we obtain the following constraint

$$D^{++}D^{--}\tilde{\xi}_k + \alpha' \partial_{++}\partial_{--}Y_{cl\,k} + \beta' \tilde{R}_{akld} D^{++}Y^a_{cl} D^{++}D^{--}\tilde{\xi}^l D^{--}Y^d_{cl} = 0 \qquad (2.45)$$

We note that for $\alpha' = \beta' = 0$ we obtain the equation of motion of the dual action $\tilde{S}^{\sigma}_{(4,4)}\big|_{Y_{cl}=0}$ which is equivalent to the equation of motion of the original action $S^{\sigma}_{(4,4)}\big|_{\Omega_{cl}=0}$. Moreover, the constraints of the (4,4) supersymmetric sigma model and its dual are equivalent by using the following equalities

$$\begin{aligned} \tilde{\xi}_a &= \{\tilde{\xi}_0 \;,\; \tilde{\xi}_i = \xi_i\} \\ Y^a_{cl} &= \{\tilde{\varphi} \;,\; \Omega^a_{cl}\} \\ \beta' \tilde{R}_{akld} &= \beta R_{akld} \end{aligned} \qquad (2.46)$$

## 3 - Quantum Equivalence of (4,4) Supersymmetric Sigma Models Related by Poisson-lie T-Duality

Let us consider the two-dimensional (4,4) supersymmetric sigma model [7,8] which is described on the target supermanifold M by a metric $G_{ab}$, a = 1, . . . , d and antisymmetric tensor $B_{ab}$

$$S^{\sigma}_{(4,4)} = \int d\mu \left\{ F_{ab}(\Omega) D^{++}\Omega^a D^{--}\Omega^b \right\} \qquad (3.1)$$

where $F_{ab} = G_{ab} + B_{ab}$. The structure group G of the target space defines a left (right) group action namely

$$\delta \Omega^a = \varepsilon^i \vartheta^a_i \qquad (3.2)$$

with i= 1, . . . , dimG, $\varepsilon^i$ are the world-sheet dependent superparameters and $\vartheta^a_i$ are the correspondingly right (left) invariant frames in the lie superalgebra **G** of the group G which satisfy the relation

$$[\vartheta_i, \vartheta_j]^a = f_{ij}^k \vartheta_k^a \tag{3.3}$$

where $f_{ij}^k$ are the structure constants of the lie group G. Furthermore, the variation of the group G element is as follows

$$\delta g = g\varepsilon \tag{3.4}$$

or equivalently

$$g^{-1}\delta g = \varepsilon \tag{3.5}$$

On the other hand, the BRST transformations associated with (3.2) are given by

$$\begin{aligned} s\Omega^a &= C^i \vartheta_i^a \\ sC^a &= 0 \end{aligned} \tag{3.6}$$

In order to define a (4,4) supersymmetric quantum action with ghost number zero we introduce a superfield $C'^a$ and an auxiliary superfield $b^a$ such that

$$\begin{aligned} sC'^a &= b^a \\ sb^a &= 0 \end{aligned} \tag{3.7}$$

Since the classical lagrangian namely

$$L_{cl} = F_{ab}(\Omega) D^{++}\Omega^a D^{--}\Omega^b$$

is invariant under the BRST transformations (3.6) then $L_{cl}$ can be replaced by

$$L_Q = L_{cl} + s\hat{L} \tag{3.8}$$

Consequently, the (4,4) supersymmetric quantum sigma model action is given by

$$S^\sigma_{(4,4)Q} = \int d\mu \left[ L_{cl} + s\hat{L} \right] \tag{3.9}$$

with

$$\hat{L} = F_{ab}(\Omega)\left[ D^{++}C'^a D^{--}\Omega^b + D^{++}\Omega^a D^{--}C'^b \right]$$

This quantum action is expanded as:

$$S^{\sigma}_{(4,4)Q} = \int d\mu \left\{ F_{ab}(\Omega) \, D^{++}\Omega^a \, D^{--}\Omega^b + C^j \left[ D^{--}(F_{ba}\vartheta^a_j D^{++}C'^a) - D^{++}(F_{ab}\vartheta^a_j D^{--}C'^b) \right] \right.$$
$$\left. C^i \left[ \partial_d F_{ab} \vartheta^d_j D^{--}C'^b D^{++}\Omega^a + \partial_d F_{ab} \vartheta^d_i D^{++}C'^a D^{--}\Omega^b \right] \right\} \quad (3.10)$$

where we have used the constraint

$$D^{++}(F_{ab} \, D^{--}\Omega^b) + D^{--}(F_{ba} \, D^{++}\Omega^b) = 0 \quad (3.11)$$

which is obtained by eliminating the auxiliary superfield $b^a$. Now let us give the variation of the action (3.10) under the BRST transformations (3.6) and (3.7) namely

$$sS^{\sigma}_{(4,4)Q} = \int d\mu \; C^j \left\{ D^{--}J^{++}_{Qj} - D^{++}J^{--}_{Qj} + \partial_d F_{ab} \, \vartheta^d_j D^{++}\Omega^a D^{--}\Omega^b + \partial_d sF_{ab} \, \vartheta^d_j D^{--}C'^b D^{++}\Omega^a \right.$$
$$+ \partial_d sF_{ab} \, \vartheta^d_j D^{++}C'^a D^{--}\Omega^b + \partial_d F_{ab} \, \vartheta^d_j D^{--}C'^b D^{++}C^i \vartheta^a_i$$
$$\left. + \partial_d F_{ab} \, \vartheta^d_j \vartheta^b_i D^{++}C'^a D^{--}C^i \right\}$$
$$\quad (3.12)$$

where the following constraint

$$D^{--} \left[ \partial_d F_{ab} \, \vartheta^d_j D^{++}\Omega^a - D^{++}C^j F_{ab} \, \vartheta^a_j \right] + D^{++} \left[ \partial_b F_{ba} \, \vartheta^d_j D^{--}\Omega^a + D^{--}C^j F_{ba} \, \vartheta^a_j \right] = 0 \quad (3.13)$$

which is deduced by eliminating the auxiliary superfield, is used and the quantum supercurrents $J^{\pm\pm}_Q$ are given by

$$J^{++}_{Qj} = F_{ba}\vartheta^a_j D^{++}\Omega^b + (sF_{ba})\vartheta^a_j D^{++}C'^b$$
$$J^{--}_{Qj} = F_{ab}\vartheta^a_j D^{--}\Omega^b + (sF_{ab})\vartheta^a_j D^{--}C'^b \quad (3.14)$$

Consequently, the BRST invariance of the quantum action (3.10) leads to the following equation

$$D^{--}J^{++}_{Qj} - D^{++}J^{--}_{Qj} + \partial_d F_{ab}\vartheta^d_j [D^{++}\Omega^a D^{--}\Omega^b + D^{--}C'^b D^{++}C^i \vartheta^a_i + \vartheta^b_i D^{++}C'^a D^{--}C^i]$$
$$+ \partial_d sF_{ab}\vartheta^d_j [D^{--}C'^b D^{++}\Omega^a + D^{++}C'^a D^{--}\Omega^b] = 0 \quad (3.15)$$

which can be rewritten as

$$(d+s)J_{Qj} + l_{\vartheta_j}(L_Q) = 0 \quad (3.16)$$

where $l_{\vartheta_j}$ denotes the lie derivative namely

$$l_{\vartheta_j} F_{ab} = \partial_d F_{ab} \vartheta_j^d$$
$$l_{\vartheta_j} sF_{ab} = \partial_d sF_{ab} \vartheta_j^d \tag{3.17}$$

d is the exterior derivative on the analytic subspace [8] and

$$J_{Qj} = J_{Qj}^{--} d\eta^{++} + J_{Qj}^{++} d\eta^{--} \tag{3.18}$$

are the Noetherian 1-forms on the world-sheet where the harmonic differentials are given by

$$d\eta^{\pm\pm} = U^{\pm\pm\alpha} dU_\alpha^{\pm\pm}$$

On the other hand the action (3.10) possesses non-commutative conservation laws if the generalized Cartan-Maurer equation holds on shell

$$(d+s)J_{Qi} + \frac{1}{2} \tilde{f}_i^{jk} J_{Qj} \wedge J_{Qk} = 0 \tag{3.19}$$

where $\tilde{f}_i^{jk}$ are the structure constants of the dual target space with lie superalgebra $\tilde{\mathbf{G}}$. Thereafter, from (3.16) and (3.19) we deduce that

$$l_{\vartheta_i}(L_Q) = \frac{1}{2} \tilde{f}_i^{jk} J_{Qj} \wedge J_{Qk} \tag{3.20}$$

Furthermore, the expansion of (3.19) in terms of ghost number leads to

$$dJ_{Qi} + \frac{1}{2} \tilde{f}_i^{jk} J_{Qj} \wedge J_{Qk} = 0$$
$$sJ_{Qi} = 0 \tag{3.21}$$

which are equivalent to

$$\partial_{\pm\pm} J_{Qi}^{\pm\pm} = 0$$
$$D_-^- J_{Qi}^{\pm\pm} = 0 = D_+^+ J_{Qi}^{\pm\pm}$$
$$sJ_{Qi}^{\pm\pm} = 0 \tag{3.22}$$
$$D^{--} J_{Qi}^{++} - D^{++} J_{Qi}^{--} + J_{Qj}^{++} \tilde{f}_i^{jk} J_{Qk}^{--} = 0$$

By using the expressions of the supercurrents (3.14), the component equations (3.22) give the following conditions for $F_{ab}$ and its BRST transformations namely

$$l_{\vartheta_i}(F_{ab}) = F_{ca}\vartheta_j^c \tilde{f}_i^{jk} F_{db}\vartheta_k^d \tag{3.23.a}$$

$$l_{\vartheta_i}(sF_{ab}) = F_{ca}\vartheta_j^c \tilde{f}_i^{jk} sF_{db}\vartheta_k^d$$
$$l_{\vartheta_i}(sF_{ab}) = sF_{ca}\vartheta_j^c \tilde{f}_i^{jk} F_{db}\vartheta_k^d \tag{3.23.b}$$

$$sF_{ea}\vartheta_j^a \tilde{f}_i^{jk} sF_{db}\vartheta_k^d D^{++}C'^e D^{--}C'^b = 0 \tag{3.24}$$

Thereafter, the equation (3.24) implies that

$$J_{C'j}^{++} \tilde{f}_i^{jk} J_{C'k}^{--} = 0$$

or equivalently

$$J_{C'j} \wedge J_{C'k} = 0 \tag{3.25}$$

The equations (3.23.a) and (3.23.b) are the conditions of the Poisson-Lie symmetry formulated at the level of the (4,4) supersymmetric sigma model quantum lagrangian, as for the ordinary situation [4,6,19,8], by adding the condition (3.23.b) given by the lie derivative of the BRST transformation of the tensor $F_{ab}(\Omega)$. However, we conclude that the (4,4) supersymmetric quantum sigma model with the action of the group G on its target space admits a Poisson-Lie dual model for some dual group $\tilde{G}$ [6,20]. Therewith, the dual quantum action of (3.1) is given by

$$\tilde{S}_{(4,4)Q}^{\sigma} = \int d\mu \left\{ \tilde{F}^{ab}(\tilde{\Omega}) D^{++}\tilde{\Omega}_a D^{--}\tilde{\Omega}_b + s\left[\tilde{F}_{ab}(\tilde{\Omega})(D^{++}\tilde{C}'^a D^{--}\tilde{\Omega}^b + D^{++}\tilde{\Omega}^a D^{--}\tilde{C}'^b\right]\right\} \tag{3.26}$$

which satisfies the following conditions

$$l_{\tilde{\vartheta}^i}(\tilde{F}^{ab}) = \tilde{F}^{ca}\tilde{\vartheta}_c^j f_{jk}^i \tilde{\vartheta}_d^k \tilde{F}^{db}$$
$$l_{\tilde{\vartheta}^i}(s\tilde{F}^{ab}) = \tilde{F}^{ca}\tilde{\vartheta}_c^j f_{jk}^i \tilde{\vartheta}_d^k s\tilde{F}^{db} = s\tilde{F}^{ca}\tilde{\vartheta}_c^j f_{jk}^i \tilde{\vartheta}_j^k \tilde{F}^{db} \tag{3.27}$$

and where the backgrounds are related by

$$[F_{ab}(\Omega = 0)]^{-1} = \tilde{F}^{ab}(\tilde{\Omega} = 0)$$
$$[sF_{ab}(\Omega = 0)]^{-1} = \left[s(\tilde{F}^{ab}(\tilde{\Omega} = 0))^{-1}\right]^{-1} \tag{3.28}$$

Finally, we note that our results on Poisson-Lie T-duality and especially our general formulation of T-duality in non isometric backgrounds may applied to a wide class of non linear sigma models [21]. On the other hand, supersymmetric quantum cosmologies may be derived from the non-linear sigma model with appropriate linearity conditions [22]. Therefore, it would be interesting to investigate N = 4 supersymmetric quantum cosmologies from the (4,4) supersymmetric non-linear sigma models associated with duality [23].

## 4 - Conclusion

In this paper, we have constructed the quantum actions of the (4,4) supersymmetric non-linear sigma model and its dual in the Abelian case by using the background superfield method, which is based on the parallel transport equation. Furthermore, we have deduced the relations between the curvature tensor of the supermanifold M and its dual by using the Buscher's formulas. The propagators of the quantum superfields and its dual are determined by introducing the n-bein $e_i^a(\Omega_{cl})$ and its dual $\tilde{e}_i^a(Y_{cl})$.

On the other hand, by using the BRST transformations associated with the left (right) group action on the superfields of the supermanifold M, we have constructed the quantum (4,4) supersymmetric non-Abelian dual sigma model action. This is obtained in the sense of Poisson-Lie T-duality which generalizes the Abelian and non-Abelian dualities. The quantum action and its dual obey the same conditions of the Poisson-Lie symmetry but with the tilted and untilted variables interchanged. Thus, the non-commutative conservation laws for the quantum (4,4) supersymmetric sigma model are given in terms of the quantum supercurrents $J_Q^{\pm\pm}$. However, the investigation of the N = 4 supersymmetric quantum cosmologies from the (4,4) supersymmetric non-linear sigma models associated with duality is under study [23].

## Acknowledgments

The authors would like to thank Professor S. Randjbar-Deami for reading the manuscript and Professor M. Virasoro, the International Atomic Energy Agency and UNESCO for hospitality at the Abdus Salam International Centre for Theoretical Physics, Trieste. This work is supported by the program PARS n$^0$ phys. 27.372/98 CNR and by the frame work of the Associate and Federation Schemes of the Abdus Salam International Centre for Theoretical Physics, Trieste.15

**References**

[1]  T. Buscher, Phys. Lett. B**194**, 51 (1987); E. Alvarez, L. Alvarez-Gaume and Y. Lozano, Nucl. Phys. (Proc. Suppl) B**41**, 1 (1995); A. Giveon, M. Porati, E. Rabinovici, Phys. Rep**244**, 77 (1994).

[2]  X. De la Ossa and F. Quevedo, Nucl. Phys. B**403**, 377 (1993); B. E. Fridling and A. Jevicki, Phys. Lett. B**134**, 70 (1984); E. S. Fradkin, A. A. Tseytlin, Ann. Phys. **162**, 31 (1985); M. J. O'Loughlim and S. Randjbar-Deami, ICTP prepeint $N^0$ IC 98100.

[3]  T. Curtright and C. Zachos, Phys. Rev. D**49**, 5408 (1994); Phys. Rev. D**52**, 573 (1995).

[4]  C. Klimcik, P. Severa, Phys. Lett. B**351**, 455 (1995); C. Klimcik, Nucl. Phys. (Proc. Suppl) B**46**, 116 (1996).

[5]  V. G. Drinfeld, Quantum Groups, Proc. Int. Cong. Math, Berkley, Calif 798 (1986).

[6]  C. Klimcik, P. Severa, Phys. Lett. B**372**, 65 (1996); Phys. Lett. B**383**, 281 (1996); "T-duality and the moment map", IHES/P/96/70 hep-th/9610198; E. Witten, J. Geom. Phys. **22**, 1 (1997); E. Tyurin, R. Von. Unge, Phys. Lett. B**382**, 233 (1996).

[7]  F. Assaoui, N. Benhamou, T. Lhallabi, Int. J. Mod. Phys. A, Vol. 14, No. 32, (1999) 5093-5104.

[8]  F. Assaoui, T. Lhallabi, Lab/UFR-HEP/00/04, to appear in Int. Jour. Mod. Phys. A.

[9]  J. Balog, P. Forgacs, N. Mohammedi, L. Palla, J. Schnittger, Nucl. Phys. B**353**, 461 (1998).

[10]  K. Sfetsos, Phys. Lett. B4**32**, 365 (1998).

[11]  L.Alvarez-Gaume, D. Z. Freedman, S. Mukhi, Ann. Phys **134**, 85 (1981); L.Alvarez-Gaume, Nucl. Phys. B**184**, 180 (1981); S. Mukhi, Nucl. Phys. B**264**, 640 (1986).

[12]  T. Lhallabi, Class. Quantum. Grav. **6**, 883 (1989).

[13]  S. J. Jr. Gates, C. M. Hull, M. Recek, Nucl. Phys. B**248**, 157 (1984); T. L. Curtright, C. K. Zachos, Phys. Rev. Lett. **53**, 1799 (1984).

[14]  T. Lhallabi, E. H Saidi, Int.J. Mod. Phys.A**3**, 187 (1988); Int.J. Mod. Phys.A**3**, 419 (1988).

[15]  P. S. Howe, G. J. Papadopoulos, K. S. Stelle, PRE 29754-Nov 1986, 24p, Preprint Imprial /TP/85-86/13.

[16]  J. Honerkamp, Nucl. Phys. B**36**, 130 (1972).

[17]  A. Galperin, E. Ivanov, V. Ogievetsky, E. Sokatchev, Class. Quantum. Grav. **2**, 601 (1985).

[18]  J. Balog, P. Forgacs, Z. Horvath, L. Palla, Phys. Lett. B**388**, 121 (1996).

[19]  E. Tyurin, Phys. Lett. B**364**, 157 (1994).

[20]  K. Sfetsos, Nucl. Phys. (Pro. Suppl). B**56**, 302 (1997); A. Alekssev, A. Malkin, Com. Math. Phys. **162**, 147 (1994).

[21]  F. Assaoui, N. Benhamou, T. Lhallabi, In Preparation.

[22]  J. E. Lidsey, J. Maharana, Phys. Rev. D**,** CERN-TH/98-022, gr. qc/9801090.

[23]  F. Assaoui, T. Lhallabi, In Preparation.